\font\tenfrakturb=eufb10
\font\tenfraktur=eufm10
\font\tenmsbm=msbm10
\font\sevenfrakturb=eufb7
\font\sevenfraktur=eufm7
\font\sevenmsbm=msbm7
\font\fivefrakturb=eufb5
\font\fivefraktur=eufm5
\font\fivemsbm=msbm5
\newfam\bgothicfam
\newfam\gothicfam
\newfam\msbmfam
\textfont\bgothicfam = \tenfrakturb \scriptfont\bgothicfam=\sevenfrakturb
\scriptscriptfont\bgothicfam=\fivefrakturb
\textfont\gothicfam = \tenfraktur \scriptfont\gothicfam=\sevenfraktur
\scriptscriptfont\gothicfam=\fivefraktur
\textfont\msbmfam = \tenmsbm \scriptfont\msbmfam=\sevenmsbm
\scriptscriptfont\msbmfam=\fivemsbm

\def\Bbb{\tenmsbm\fam\msbmfam}

\catcode`@=11
\def\renewcounter#1{\@definecounter{#1}\@ifnextchar[{\@newctr{#1}}{}}

\documentstyle[twoside]{article}

\catcode`\@=11
\long\def\@makefntext#1{
\protect\noindent \hbox to 3.2pt {\hskip-.9pt
$^{{\eightrm\@thefnmark}}$\hfil}#1\hfill}		%CAN BE USED

\def\@makefnmark{\hbox to 0pt{$^{\@thefnmark}$\hss}}	%ORIGINAL

\def\ps@myheadings{\let\@mkboth\@gobbletwo
\def\@oddhead{\hbox{}
\rightmark\hfil\eightrm\thepage}
\def\@oddfoot{}\def\@evenhead{\eightrm\thepage\hfil
\leftmark\hbox{}}\def\@evenfoot{}
\def\sectionmark##1{}\def\subsectionmark##1{}}

\oddsidemargin=\evensidemargin
\newcounter{sectionc}\newcounter{subsectionc}\newcounter{subsubsectionc}
\renewcommand{\section}[1] {\vspace{12pt}\addtocounter{sectionc}{1}
\setcounter{subsectionc}{0}\setcounter{subsubsectionc}{0}\noindent
	{\tenbf\thesectionc. #1}\par\vspace{5pt}}
\renewcommand{\subsection}[1] {\vspace{12pt}\addtocounter{subsectionc}{1}
	\setcounter{subsubsectionc}{0}\noindent
	{\bf\thesectionc.\thesubsectionc. {\kern1pt \bfit #1}}\par\vspace{5pt}}
\renewcommand{\subsubsection}[1] {\vspace{12pt}\addtocounter{subsubsectionc}{1}
	\noindent{\tenrm\thesectionc.\thesubsectionc.\thesubsubsectionc.
	{\kern1pt \tenit #1}}\par\vspace{5pt}}
\newcommand{\nonumsection}[1] {\vspace{12pt}\noindent{\tenbf #1}
	\par\vspace{5pt}}

\newcounter{appendixc}
\newcounter{subappendixc}[appendixc]
\newcounter{subsubappendixc}[subappendixc]
\renewcommand{\thesubappendixc}{\Alph{appendixc}.\arabic{subappendixc}}
\renewcommand{\thesubsubappendixc}
	{\Alph{appendixc}.\arabic{subappendixc}.\arabic{subsubappendixc}}

\renewcommand{\appendix}[1] {\vspace{12pt}
        \refstepcounter{appendixc}
        \setcounter{figure}{0}
        \setcounter{table}{0}
        \setcounter{lemma}{0}
        \setcounter{theorem}{0}
        \setcounter{corollary}{0}
        \setcounter{definition}{0}
        \setcounter{equation}{0}
        \renewcommand{\thefigure}{\Alph{appendixc}.\arabic{figure}}
        \renewcommand{\thetable}{\Alph{appendixc}.\arabic{table}}
        \renewcommand{\theappendixc}{\Alph{appendixc}}
        \renewcommand{\thelemma}{\Alph{appendixc}.\arabic{lemma}}
        \renewcommand{\thetheorem}{\Alph{appendixc}.\arabic{theorem}}
        \renewcommand{\thedefinition}{\Alph{appendixc}.\arabic{definition}}
        \renewcommand{\thecorollary}{\Alph{appendixc}.\arabic{corollary}}
        \renewcommand{\theequation}{\Alph{appendixc}.\arabic{equation}}
        \noindent{\tenbf Appendix \theappendixc #1}\par\vspace{5pt}}
\newcommand{\subappendix}[1] {\vspace{12pt}
        \refstepcounter{subappendixc}
        \noindent{\bf Appendix \thesubappendixc. {\kern1pt \bfit #1}}
	\par\vspace{5pt}}
\newcommand{\subsubappendix}[1] {\vspace{12pt}
        \refstepcounter{subsubappendixc}
        \noindent{\rm Appendix \thesubsubappendixc. {\kern1pt \tenit #1}}
	\par\vspace{5pt}}

\newcommand{\textlineskip}{\baselineskip=13pt}
\newcommand{\smalllineskip}{\baselineskip=10pt}

\def\eightcirc{
\begin{picture}(0,0)
\put(4.4,1.8){\circle{6.5}}
\end{picture}}
\def\eightcopyright{\eightcirc\kern2.7pt\hbox{\eightrm c}}

\newcommand{\copyrightheading}[1]
	{\vspace*{-2.5cm}\smalllineskip{\flushleft
	{\footnotesize International Journal of Modern Physics D
          #1}\\
	{\footnotesize $\eightcopyright$\, World Scientific Publishing
	 Company}\\
	 }}

\newcommand{\pub}[1]{{\begin{center}\footnotesize\smalllineskip
	Received #1\\
	\end{center}
	}}

\def\abstracts#1#2#3{{
	\centering{\begin{minipage}{4.5in}\baselineskip=10pt\footnotesize
	\parindent=0pt #1\par
	\parindent=15pt #2\par
	\parindent=15pt #3
	\end{minipage}}\par}}

\newcommand{\bibit}{\nineit}
\newcommand{\bibbf}{\ninebf}
\renewenvironment{thebibliography}[1]
        {\frenchspacing
	 \ninerm\baselineskip=11pt
         \begin{list}{\arabic{enumi}.}
        {\usecounter{enumi}\setlength{\parsep}{0pt}
	 \setlength{\leftmargin 12.7pt}{\rightmargin 0pt} %FOR 1--9 ITEMS
         \setlength{\itemsep}{0pt} \settowidth
	{\labelwidth}{#1.}\sloppy}}{\end{list}}

\newcounter{itemlistc}
\newcounter{romanlistc}
\newcounter{alphlistc}
\newcounter{arabiclistc}

\newcommand{\fcaption}[1]{
        \refstepcounter{figure}
        \setbox\@tempboxa = \hbox{\footnotesize Fig.~\thefigure. #1}
        \ifdim \wd\@tempboxa > 5in
           {\begin{center}
        \parbox{5in}{\footnotesize\smalllineskip Fig.~\thefigure. #1}
            \end{center}}
        \else
             {\begin{center}
             {\footnotesize Fig.~\thefigure. #1}
              \end{center}}
        \fi}

\newcommand{\tcaption}[1]{
        \refstepcounter{table}
        \setbox\@tempboxa = \hbox{\footnotesize Table~\thetable. #1}
        \ifdim \wd\@tempboxa > 5in
           {\begin{center}
        \parbox{5in}{\footnotesize\smalllineskip Table~\thetable. #1}
            \end{center}}
        \else
             {\begin{center}
             {\footnotesize Table~\thetable. #1}
              \end{center}}
        \fi}

\def\@citex[#1]#2{\if@filesw\immediate\write\@auxout
	{\string\citation{#2}}\fi
\def\@citea{}\@cite{\@for\@citeb:=#2\do
	{\@citea\def\@citea{,}\@ifundefined
	{b@\@citeb}{{\bf ?}\@warning
	{Citation `\@citeb' on page \thepage \space undefined}}
	{\csname b@\@citeb\endcsname}}}{#1}}

\newif\if@cghi
\def\cite{\@cghitrue\@ifnextchar [{\@tempswatrue
	\@citex}{\@tempswafalse\@citex[]}}
\def\citelow{\@cghifalse\@ifnextchar [{\@tempswatrue
	\@citex}{\@tempswafalse\@citex[]}}
\def\@cite#1#2{{$\null^{#1}$\if@tempswa\typeout
	{IJCGA warning: optional citation argument
	ignored: `#2'} \fi}}

\def\pmb#1{\setbox0=\hbox{#1}
	\kern-.025em\copy0\kern-\wd0
	\kern.05em\copy0\kern-\wd0
	\kern-.025em\raise.0433em\box0}

\def\fnt#1#2{\footnotetext{\kern-.3em
	{$^{\mbox{\scriptsize #1}}$}{#2}}}

\def\fpage#1{\begingroup
\voffset=.3in
\thispagestyle{empty}\begin{table}[b]\centerline{\footnotesize #1}
	\end{table}\endgroup}

\def\runninghead#1#2{\pagestyle{myheadings}
\markboth{{\protect\footnotesize\it{\quad #1}}\hfill}
{\hfill{\protect\footnotesize\it{#2\quad}}}}
\font\tenrm=cmr10
\font\tenit=cmti10
\font\tenbf=cmbx10
\font\bfit=cmbxti10 at 10pt
\font\ninerm=cmr9
\font\nineit=cmti9
\font\ninebf=cmbx9
\font\eightrm=cmr8

\def\ghc{{\sqrt G\over \hbar c}}
\def\qed{\hbox{${\vcenter{\vbox{  %HOLLOW SQUARE
   \hrule height 0.4pt\hbox{\vrule width 0.4pt height 6pt
   \kern5pt\vrule width 0.4pt}\hrule height 0.4pt}}}$}}

\begin{document}
\runninghead{N. E. Firsova}
{Correct Statement of a Scattering Problem for Quantum Charged Scalar
Particles on the...}

\normalsize\textlineskip
\thispagestyle{empty}
\setcounter{page}{1}
\copyrightheading{}
\vspace*{0.88truein}
\fpage{1}
\centerline{\bf CORRECT STATEMENT OF A SCATTERING PROBLEM }
\vspace*{0.035truein}
\centerline{\bf  FOR QUANTUM CHARGED SCALAR PARTICLES}
\vspace*{0.035truein}
\centerline{\bf ON THE REISSNER-NORDSTR\"{O}M BLACK HOLES
}
\vspace*{0.37truein}
\centerline{\footnotesize N. E. FIRSOVA}
\vspace*{0.015truein}
\centerline{\footnotesize\it Institute for Mechanical Engineering,
Russian Academy of Sciences}
\baselineskip=10pt
\centerline{\footnotesize\it Sankt-Petersburg 199178, Russia}
\vspace*{0.225truein}
\pub{: August 1997}
\vspace*{0.21truein}
\abstracts{
 We study a correct statement of the scattering problem arising for quantum
charged scalar particles on the Reissner-Nordstr\"{o}m black holes when
taking into account the own electric field of black hole. The elements of the
corresponding $S$-matrix are explored in the form convenient to physical
applications and for applying numerical methods. Some further possible
issues are outlined.
}{}{}
%\vspace*{10pt}
%\keywords{The contents of the keywords}
%\textlineskip  %) USE THIS MEASUREMENT WHEN THERE IS
%\vspace*{12pt}	%) NO SECTION HEADING
\vspace*{1pt}\textlineskip %) USE THIS MEASUREMENT WHEN THERE IS
\section{Introduction} %) A SECTION HEADING
\vspace*{-0.5pt}
\noindent

 As is well known (see, e. g., Ref.\cite
{NF86}), when calculating
miscellaneous quantum effects in the 4D black hole physics one encounters
many scattering problems. These problems can be split into two
classes: the ones on the semiaxis and the ones on the axis. The former
class is important when analysing, for example, the processes of the
particle absorption by black holes. The latter one, to our mind, more
important for the 4D black hole physics, is connected with semiclassical
quantization of various physical fields on the background of the black holes
and with quantum effects arising within the framework of this approach, for
instance, such as the Hawking radiation process. Under the circumstances the
situation with a more or less strict study of the mentioned problems should,
however, be considered as unsatisfactory.

As a rule, the investigators treat
the given problems at a qualitative level using one or another approximation
without any justification from mathematical point of view. But many effects
in the 4D black hole physics can be computed only numerically and when doing
so it is very important to know whether the quantities under evaluation exist
in the strict mathematical sense. The given quantities, however, often contain
the characteristics related with the mentioned scattering problems, for
example, the elements of the corresponding $S$-matrices.
It is evident that
in this situation the physical value of such computations is extremely
low and they can serve only as a crude approximation.

The present note is intended to consider correctly one of the abovementioned
scattering problems which arises when studying topologically inequivalent
configurations (TICs) of complex scalar field on the 4D black holes. As was
discussed in Refs.\cite{Gon94}, the standard spacetime topology
${\Bbb{R}}^2\times S^2$ of the 4D black hole physics admits the countable
number of TICs for the complex scalar field, each TIC being labeled by its
Chern number $n\in{\Bbb{Z}}$. In its turn, this yields, for instance, the
additional contributions to the Hawking radiation due to the twisted TICs,
i. e., the ones with $n\ne0$. From physical point of view such contributions
are conditioned by the natural presence of Dirac magnetic $U(1)$-monopoles
on black holes. This question was explored in
Refs.\cite{{GF96},{GF97}} for the Scwarzschild (S) and Reissner-Nordstr\"{o}m
(RN) black holes. Referring for more details to Refs.\cite{{GF96},{GF97}},
we shall here adduce the expression for the luminosity $L(n,\alpha)$ with
respect to the Hawking radiation for TIC with the Chern number $n$ for the
case S and RN black holes. It looks as follows
\begin{equation}              \label{1.1a}
 L(n,\alpha)=2A\sum\limits_{l=|n|}^\infty(2l+1)\,
\int\limits_0^\infty B\frac{|s_{11}(k,\alpha,l)|^2\,k\,dk}
{e^{ 4\pi k[1+f(\alpha)]}-B}
\end{equation}
with dimensionless $k$ and
$f(\alpha)={1\over2}(\sqrt{1-\alpha^2}+1/\sqrt{1-\alpha^2})$, while
$A={1\over2\pi\hbar}\left({\hbar c^3\over GM}\right)^2\approx
0.273673\cdot10^{50}\,{\rm{erg\cdot s^{-1}}}\cdot M^{-2}$ ($M$ in g),
$e=4.8\cdot10^{-10}\,{\rm cm^{3/2}\cdot g^{1/2}\cdot s^{-1}}$,
$\alpha=|Q|/M$ with $0\leq\alpha<1$, $M$, $Q$ are, respectively,
a black hole mass and an electric charge,
and $B=\exp[-4\pi eQ\ghc{1+f(\alpha)\over1+\sqrt{1-\alpha^2}}]$.

     In formula (\ref{1.1a}) we assume that $eQ>0$ or else it should be
regularized because then there is a singularity in the integral.

To find the coefficient $s_{11}$, one should consider a scattering problem
on the axis
for the Schr\"{o}dinger type equation
\begin{equation}                      \label{1.1}
[-\frac{d^2}{dx^2}+q(x,\alpha,l)]\psi(x,k,\alpha,l)=
(k+\frac{eQ}{y(x)})^2\psi(x,k,\alpha,l),
\end{equation}
\noindent
\begin{equation}                  \label{1.2}
q(x,\alpha,l)=\left[1-\frac{2}{y(x)}+\frac{\alpha^2}{y^2(x)}\right]
\left[\frac{N}{y^2(x)}+\frac{2}{y^3(x)}-\frac{2\alpha^2}{y^4(x)}\right],
\end{equation}
\noindent
where $y(x)$ is a function reverse to the following one
\begin{equation}           \label{1.3}
x(y)=y+\frac{y_+^2}{y_+-y_-}\ln\left|\frac{y-y_+}{2}\right| -
\frac{y_-^2}{y_+-y_-}
\ln\left|\frac{y-y_-}{2}\right|,
\end{equation}
\begin{equation}          \label{1.4}
y_{\pm}=1\pm\sqrt{1-\alpha^2},
\end{equation}
\noindent
so $y(x)$ is the one-to-one correspondence between ($-\infty,\infty$) and
($y_+,\infty$). Besides, $N=l(l+1)$ or $N=l(l+1)-n^2$ (in both the cases
$l=|n|, |n|+1,...$) in dependence of the gauge choice for connection $A_\mu$
(vector-potential for the conforming Dirac monopole) in the line bundle with
the Chern number $n$. Both the gauges are interesting from physical point of
view but for the considerations in the present paper the concrete choice is
inessential, therefore, in what follows, the notation $q(x,\alpha,l)$ can
be taken as the common one for both the cases.

     In $4D$ black holes physics literature (see, e. g., Ref.\cite{NF86}) it is
proposed to determine the scattering functions as the solutions of
the equation (\ref{1.1}) which obey the following conditions
\begin{equation}    \label{1.1b}
\Psi^1(x,k,\alpha,l)\sim\cases{e^{i(k+eQ/y_+)x}+
s_{12}(k,\alpha,l)e^{-i(k+eQ/y_+)x}\>,
&$x\to -\infty$,\cr
s_{11}(k,\alpha,l)e^{ikx}, &$x\to+\infty$,\cr}
\end{equation}
\begin{equation}   \label{1.1c}
\Psi^2(x,k,\alpha,l)\sim\cases{s_{22}(k,\alpha,l)
e^{-i(k+eQ/y_+)x},
&$x\to-\infty$,\cr
e^{-ikx}+s_{21}(k,\alpha,l)e^{ikx}, &$x\to+\infty$\cr}
\end{equation}
\noindent
with $S$-matrix
\begin{equation}    \label{1.1d}
S(k,\alpha,l)=\pmatrix{s_{11}(k,\alpha,l)
&s_{12}(k,\alpha,l)\cr
s_{21}(k,\alpha,l)&s_{22}(k,\alpha,l)\cr}\>.
\end{equation}
In the case of the S black hole, one should take $\alpha=Q=0$ and we come to
the standard scattering problem on the axis for the Schr\"odinger equation
which is well described in literature (see, e. g., Ref.\cite{CS77}) and it was
used in Ref.\cite{GF96}. The next important case is the RN black hole
when neglecting the influence of its own (external) electric field on the
charged scalar particles leaving the black hole. In this situation, one
should take 
$e=0$ in (\ref{1.1}), (\ref{1.1b}), (\ref{1.1c}), so
that $B=1$ in (\ref{1.1a}), and
we again come to the standard scattering problem for the Schr\"odinger
equation but with the more complicated potentilal (\ref{1.2}), where 
$\alpha\ne0$. This case was analysed in Ref.\cite{GF97}.
It is clear that the most general case with taking into account
the own electric
field of RN black hole requires considering the scattering problem
for (\ref{1.1})
at arbitrary $e\ne0$, $Q\ne0$, $\alpha\ne0$. This note is
devoted just to this case.
It is clear that
there are difficulties in this problem since
the potential $q(x,\alpha,l)$ of (\ref{1.2}) is given in an inexplicit
form.
Besides, as will be seen below,
the above statement of the scattering problem (see
(\ref{1.1}), (\ref{1.1b})--(\ref{1.1c}))
is not quite correct at $e\ne0$, $Q\ne0$, $\alpha\ne0$
and needs some changes. The reason is that the problem
(\ref{1.1}), (\ref{1.1b})--(\ref{1.1c}) has no solution
since the effective potential of the problem is equal to
$$q(x,\alpha,l)-2k\frac{eQ}{y(x)}-\frac{e^2Q^2}{y^2(x)}$$
and it is not integrable
at $x\to+\infty$ (long range potential, see below), though the
potential $q(x,\alpha,l)$ is integrable \cite{GF97}.
Therefore the problem should be regularized.

We start in Sec. 2 by obtaining the necessary estimates needed for the further
correct statement of scattering problem under consideration. In turn, this
allows us to formulate the mentioned problem in Sec. 3 while Sec. 4 contains
the study of some asymptotic properties of the conforming $S$-matrix that
can be useful, for example, for checking numerical calculations. Finally,
Sec. 5 is devoted to concluding remarks.

Physical issues of the results obtained here will be discussed elsewhere,
as they require numerical calculations like those in Refs.
\cite{{GF96},{GF97}} and the results of the present paper give the necessary
basis for these computations.
\vskip1.0cm
\section{Preliminaries}

     To study equation (\ref{1.1}) we should first explore functions
$y(x)$ and $q(x,\alpha,l)$. From (\ref{1.3}), (\ref{1.4}) it follows
that if $y$ varies from $y_+=1+\sqrt{1-\alpha^2}$ to $+\infty$
then $x$ varies from $-\infty$ to $+\infty$. We see that
$$x_{y\to y_+}=\frac{y_+^2}{2\sqrt{1-\alpha^2}}\ln |y-y_+|+O(1)\to-\infty\>$$
So we have
\begin{equation}
y(x)_{x\to-\infty} =y_++O(\exp(\frac{2\sqrt{1-\alpha^2}}{y_+^2}x))\>.
\label{2.3}
\end{equation}
Taking into consideration that $y_+^2-2y_++\alpha^2=0$ we get
$$(1-\frac{2}{y(x)}+\frac{\alpha^2}{y^2(x)})_{x\to-\infty}=
O(\exp(\frac{2\sqrt{1-\alpha^2}}{y_+^2}x))\>,$$
and (see (\ref{1.2}))
\begin{equation}
q(x,\alpha,l)_{x\to-\infty}=O(\exp(\frac{2\sqrt{1-\alpha^2}}{y_+^2}x))\>.
\label{2.5}
\end{equation}
It is clear that
\begin{equation}\label{2.2}
y(x)_{x\to+\infty}=x(1+O(\frac{\ln x}{x}))\>.
\end{equation}
\noindent
From (\ref{1.2}) and (\ref{2.2}), (\ref{2.3}) we see
\begin{equation}
q(x,\alpha,l)_{x\to+\infty}=O(x^{-2})\>.
\label{2.4}
\end{equation}
From (\ref{2.3}), (\ref{2.5}) it follows that as $x\to-\infty$ the
equation (\ref{1.1}) has the form
\begin{equation}
[\frac{d^2}{dx^2}+(k+\frac{eQ}{y_+})^2]\psi(x,k,\alpha,l)=
q^-(x,k,\alpha,l)\psi(x,k,\alpha,l)\>,
\label{2.6}
\end{equation}
where
\begin{equation}
q^-(x,k,\alpha,l)=u^-(x,k,\alpha,l)+2(k+\frac{eQ}{y_+})v^-(x,k,\alpha,l)\>,
\label{2.6a}
\end{equation}
\begin{equation}
u^-(x,k,\alpha,l)=q(x,k,\alpha,l)-v^-(x,k,\alpha,l)^2\>,
\label{2.6b}
\end{equation}
\begin{equation}
v^-(x,k,\alpha,l)=eQ(\frac{1}{y_+}-\frac{1}{y(x)})
\label{2.6c}\>.
\end{equation}
Here
$$u^-(x,k,\alpha,l)_{x\to-\infty}=
O(\exp(\frac{2\sqrt{1-\alpha^2}}{y_+^2}x)),$$
$$v^-(x,k,\alpha,l)_{x\to-\infty}=
O(\exp(\frac{2\sqrt{1-\alpha^2}}{y_+^2}x)),$$
and consequently
\begin{equation}
\int^a_{-\infty}|u^-(x,k,\alpha,l)|dx<C(a),
\int^a_{-\infty}|v^-(x,k,\alpha,l)|dx<C(a).
\label{2.7a}
\end{equation}
\noindent
From (\ref{2.2}), (\ref{2.4}) it follows that as $x\to+\infty$ the equation
(\ref{1.1}) has the form
\begin{equation}
[\frac{d^2}{dx^2}+(k^2+2k\frac{eQ}{x})]\psi(x,k,\alpha,l)=
q^+(x,k,\alpha,l)\psi(x,k,\alpha,l),
\label{2.8}
\end{equation}
where
\begin{equation}
q^+(x,k,\alpha,l)=u^+(x,k,\alpha,l)+2kv^+(x,k,\alpha,l),
\label{2.9}
\end{equation}
\begin{equation}
u^+(x,k,\alpha,l)=q(x,k,\alpha,l)- \frac{e^2Q^2}{y^2(x)}\>,
\label{2.9b}
\end{equation}
\begin{equation}
v^+(x,k,\alpha,l)=eQ(\frac{1}{x}-\frac{1}{y(x)})\>.
\label{2.9c}
\end{equation}
Hence, as $x\to+\infty$ (see (\ref{2.2}), (\ref{2.4}))
$$u^+(x,k,\alpha,l)=O(x^{-2}), \qquad v^+(x,k,\alpha,l)=O(x^{-2})\>.$$
As a result
\begin{equation}
\int_a^{+\infty}|u^+(x,k,\alpha,l)|dx<c(a),
\int_a^{+\infty}|v^+(x,k,\alpha,l)|dx<c(a),
\label{2.9a}
\end{equation}
\noindent
The homogenous equation for (\ref{2.6}) is
\begin{equation}
[\frac{d^2}{dx^2}+(k+\frac{eQ}{y_+})^2]\psi_0^-(x,k,\alpha,l)=0,
\label{2.10}
\end{equation}
\noindent
and its general solution has the form
\begin{equation}
\psi_0^-(x,k,\alpha,l)=C_1^-e^{i(k+\frac{eQ}{y_+})x}+
C_2^-e^{-i(k+\frac{eQ}{y_+})x}.
\label{2.11}
\end{equation}
     Let us now consider the homogenous equation for (\ref{2.8})
\begin{equation}                      \label{2.12}
[\frac{d^2}{dx^2}+(k^2+2k\frac{eQ}{x})]\psi_0^+(x,k,\alpha,l)=0.
\end{equation}
\noindent
Transforming it we obtain for
$$u(z,k,\alpha,l))=\psi_0^+(\frac{z}{2ik},k,\alpha,l))$$
the Whittaker equation \cite{WW}
\begin{equation}                      \label{2.13}
u''_{zz}+\{-\frac{1}{4}-\frac{ieQ}{z}\}u=0.
\end{equation}
\noindent
A couple of the Whittaker functions $W_{-ieQ,\frac{1}{2}}(z)$,
$W_{ieQ,\frac{1}{2}}(-z)$ forms a fundamental system of solutions for
the equation (\ref{2.13}) because they have asymptotic behavior
as follows \cite{WW}
\begin{equation}                      \label{2.14}
W_{-ieQ,\frac{1}{2}}(z)=e^{\frac{1}{2}z}z^{-ieQ}(1+O(\frac{1}{z})),
|\arg z|<\pi, z\to\infty,
\end{equation}
\begin{equation}                      \label{2.15}
W_{ieQ,\frac{1}{2}}(-z)=e^{\frac{1}{2}z}(-z)^{ieQ}(1+O(\frac{1}{z})),
|\arg (-z)|<\pi, z\to\infty.
\end{equation}
\noindent
Hence the general solution of (\ref{2.13}) is
\begin{equation}                      \label{2.16}
u(z,k,\alpha,l)=
C_1^+W_{-ieQ,\frac{1}{2}}(z)+C_2^+W_{ieQ,\frac{1}{2}}(-z).
\end{equation}
\noindent
It means that the general solution of (\ref{2.12}) has the form
\begin{equation}                      \label{2.16a}
\psi_0^+(x,k,\alpha,l)=C_1^+W_{ieQ,\frac{1}{2}}(-2ikx)+
C_2^+W_{-ieQ,\frac{1}{2}}(2ikx)
\end{equation}
\noindent
and
\begin{equation}    \label{2.18}
W_{ieQ,\frac{1}{2}}(-2ikx)=
e^{ikx}|2kx|^{ieQ}e^{\frac{\pi eQ}{2}}
(1+O(\frac{1}{|kx|})), x\to\infty,
\end{equation}
\begin{equation}  \label{2.19}
W_{-ieQ,\frac{1}{2}}(2ikx)=e^{-ikx}\frac{1}{|2kx|^{ieQ}}
e^{\frac{\pi eQ}{2}}(1+O(\frac{1}{|kx|})), x\to\infty.
\end{equation}
\noindent
We introduce also functions
\begin{equation}\label{2.19a}
w_{\pm ieQ,\frac{1}{2}}(\pm z)=
W_{\pm ieQ,\frac{1}{2}}(\pm z)e^{-\frac{\pi eQ}{2}},
\end{equation}
\noindent
which are more convenient. They form the fundamental system of
solutions of the equation (\ref{2.12})as well.
We may write the general solution of (\ref{2.12}) as follows
\begin{equation}                      \label{2.17}
\psi_0^+(x,k,\alpha,l)=c_1^+w_{ieQ,\frac{1}{2}}(-2ikx)+
c_2^+w_{-ieQ,\frac{1}{2}}(2ikx)
\end{equation}
and
\begin{equation}    \label{2.18d}
w_{ieQ,\frac{1}{2}}(-2ikx)=
e^{ikx}e^{ieQ \ln|2kx|}
(1+O(|kx|^{-1}), x\to\infty,
\end{equation}
\begin{equation}  \label{2.19d}
w_{-ieQ,\frac{1}{2}}(2ikx)=e^{-ikx}e^{-ieQ \ln|2kx|}
(1+O(|kx|^{-1})), x\to\infty.
\end{equation}
The functions
$w_{-ieQ,\frac{1}{2}}(z)$,
$w_{ieQ,\frac{1}{2}}(-z)$ are complex conjugate, i.e.
\begin{equation}\label{2.20}
\overline{w_{ieQ,\frac{1}{2}}(-2ikx)}=w_{-ieQ,\frac{1}{2}}(2ikx)
\end{equation}
\noindent
and their Wronskian is equal to
\begin{equation} \label{2.21}
[w_{-ieQ,\frac{1}{2}}(2ikx),w_{ieQ,\frac{1}{2}}(-2ikx)]=2ik.
\end{equation}

\section{The Scattering Problem}
     We denote $\psi^-(x,k,\alpha,l)$ the Jost type solution of the
equation (\ref{2.6}), i.e.
\begin{equation}                                 \label{3.1}
[\frac{d^2}{dx^2}+(k+\frac{eQ}{y_+})^2]\psi^-(x,k,\alpha,l)=
q^-(x,k,\alpha,l)\psi^-(x,k,\alpha,l),
\end{equation}
\begin{equation}  \label{3.2}
\psi^-(x,k,\alpha,l)=e^{-i(k+\frac{eQ}{y_+})x}+o(1), x\to-\infty.
\end{equation}
Varying constants in (\ref{2.10}), (\ref{2.11}) according to the Lagrange
method, we obtain the integral equation equivalent to the problem (\ref{3.1}),
(\ref{3.2})
\begin{eqnarray} \label{3.3}
\psi^-(x,k,\alpha,l)=e^{-i(k+\frac{eQ}{y_+})x}+\qquad\qquad\nonumber\\
\frac{1}{k+\frac{eQ}{y_+}}\int^x_{-\infty}
\sin[(k+\frac{eQ}{y_+})(x-t)]q^-(t,k,\alpha,l)\psi^-(t,k,\alpha,l)dt.
\end{eqnarray}
\noindent
The convergence of the series of approximations for this equations
follows from (\ref{2.7a}) as usual (see, e. g., Ref.\cite{CS77}).

Differentiating equation (\ref{3.3}) we obtain expression for derivative
\begin{eqnarray}    \label{3.33}
(\psi^-)'_x(x,k,\alpha,l)=
-i(k+\frac{eQ}{y_+})e^{-i(k+\frac{eQ}{y_+})x}+\nonumber\\
\int^x_{-\infty}\cos[(k+\frac{eQ}{y_+})(x-t)]q^-(t,k,\alpha,l)
\psi^-(t,k,\alpha,l)dt
\end{eqnarray}
A couple of functions $\psi^-(x,k,\alpha,l)$,
$\overline{\psi^-(x,k,\alpha,l)}$
is the fundamental system of solutions for the equation (\ref{3.1}) and
\begin{equation} \label{3.4}
[\psi^-(x,k,\alpha,l),\overline{\psi^-(x,k,\alpha,l)}]=
2i(k+\frac{eQ}{y_+}).
\end{equation}
\noindent
We denote $\psi^+(x,k,\alpha,l)$ the Jost type solution of the equation
(\ref{2.8}), i.e.
\begin{equation}  \label{3.5}
[\frac{d^2}{dx^2}+(k^2+2k\frac{eQ}{x})]\psi^+(x,k,\alpha,l)=
q^+(x,k,\alpha,l)\psi^+(x,k,\alpha,l),
\end{equation}
\begin{equation} \label{3.6}
\psi^+(x,k,\alpha,l)=w_{ieQ,\frac{1}{2}}(-2ikx)+o(1), x\to+\infty.
\end{equation}
\noindent
Varying constants in (\ref{2.12}), (\ref{2.17}) according to the Lagrange
method,
we obtain the integral equation equivalent to the problem
(\ref{3.5}), (\ref{3.6})
\begin{eqnarray} \label{3.7}
\psi^+(x,k,\alpha,l)=w_{ieQ,\frac{1}{2}}(-2ikx)+
\frac{1}{2ik}\times\qquad\qquad\qquad\nonumber\\
\int_x^{+\infty}[w_{-ieQ,\frac{1}{2}}(2ikt)
w_{ieQ,\frac{1}{2}}(-2ikx)-w_{ieQ,\frac{1}{2}}(-2ikt)
w_{-ieQ,\frac{1}{2}}(2ikx)]\times\nonumber\\
\times q^+(x,k,\alpha,l)\psi^+(t,k,\alpha,l)dt.\qquad\qquad
\end{eqnarray}
\noindent
The existence of the solution of this equation follows from (\ref{2.9a})
as usual.
Differentiating equation (\ref{3.7}) we obtain expression for derivative
\begin{eqnarray}     \label{3.34}
(\psi^+)'_x (x,k,\alpha,l)=-2ik w'_{ieQ,\frac{1}{2}}(-2ikx)-
e^{-\pi eQ}\times\qquad\qquad\nonumber\\
\int^{+\infty}_x [w_{-ieQ,\frac{1}{2}}(2ikt)
w'_{ieQ,\frac{1}{2}}(-2ikx)+w_{ieQ,\frac{1}{2}}(-2ikt)
w'_{-ieQ,\frac{1}{2}}(2ikx)]\times\qquad\nonumber\\
q^+(t,k,\alpha,l)\psi^+(t,k,\alpha,l)dt.\qquad\qquad\qquad
\end{eqnarray}
The pair of functions
$\psi^+(x,k,\alpha,l)$, $\overline{\psi^+(x,k,\alpha,l)}$
is the fundamental system of solutions for the equation (\ref{3.5}) and
\begin{equation}  \label{3.8}
[\psi^+(x,k,\alpha,l),\overline{\psi^+(x,k,\alpha,l)}]=-2ik.
\end{equation}
\noindent
We write now decomposition of the solution $\psi^+(x,k,\alpha,l)$ to the
fundamental system $\psi^-(x,k,\alpha,l)$, $\overline{\psi^-(x,k,\alpha,l)}$,
and the solution $\psi^-(x,k,\alpha,l)$
to the pair $\psi^+(x,k,\alpha,l)$, $\overline{\psi^+(x,k,\alpha,l)}$.
We obtain
\begin{equation}  \label{3.9}
\cases{\psi^-(x,k,\alpha,l)=c_{11}(k,l)\psi^+(x,k,\alpha,l)+
c_{12}(k,l)\overline{\psi^+(x,k,\alpha,l)},\cr
\psi^+(x,k,\alpha,l)=c_{21}(k,l)\overline{\psi^-(x,k,\alpha,l)}+
c_{22}(k,l)\psi^-(x,k,\alpha,l).\cr}
\end{equation}
\noindent
We call the matrix ${\bf
 C}=\{c_{ij}\}$ {\it the transition matrix} for the equation
(\ref{1.1}). Let us explore its elements. From  (\ref{3.4}),
(\ref{3.8}), (\ref{3.9}) it follows
\begin{equation}  \label{3.10}
c_{11}(k,\alpha,l)=
\frac{1}{2ik}
[\overline{\psi^+(x,k,\alpha,l)},\psi^-(x,k,\alpha,l)],
\end{equation}
\begin{equation}   \label{3.11}
c_{12}(k,\alpha,l)=
\frac{1}{2ik}[\psi^-(x,k,\alpha,l),\psi^+(x,k,\alpha,l)],
\end{equation}
\begin{equation} \label{3.12}
c_{21}(k,\alpha,l)=
\frac{1}{2i(k+\frac{eQ}{y_+})}[\psi^-(x,k,\alpha,l),\psi^+(x,k,\alpha,l)],
\end{equation}
\begin{equation}   \label{3.13}
c_{22}(k,\alpha,l)=\frac{1}{2i(k+\frac{eQ}{y_+})}[\psi_+(x,k,\alpha,l),
\overline{\psi^-(x,k,\alpha.l)}].
\end{equation}
Let us investigate the properties of elements $c_{ij}$. For this purpose
we put
equations (\ref{3.9}) one into another. Then we have
\begin{equation}   \label{3.14}
c_{11}(k,\alpha,l)c_{22}(k,\alpha,l)+c_{12}(k,\alpha,l)
\overline{c_{21}(k,\alpha,l)}=1,
\end{equation}
\begin{equation} \label{3.15}
c_{22}(k,\alpha,l)c_{11}(k,\alpha,l)+c_{21}(k,\alpha,l)
\overline{c_{12}(k,\alpha,l)}=1,
\end{equation}
\begin{equation}\label{3.16}
c_{11}(k,\alpha,l)c_{21}(k,\alpha,l)+c_{12}(k,\alpha,l)
c_{22}(k,\alpha,l)=0,
\end{equation}
\begin{equation}\label{3.17}
c_{21}(k,\alpha,l)\overline{c_{11}(k,\alpha,l)}+
c_{22}(k,\alpha,l)c_{12}(k,\alpha,l)=0.
\end{equation}
\noindent
It follows from equations (\ref{3.10})-(\ref{3.13})
\begin{equation}\label{3.18}
\gamma(k,\alpha) c_{11}(k,\alpha,l)=-\overline{c_{22}(k,\alpha,l)},
\end{equation}
\begin{equation}\label{3.19}
\gamma(k,\alpha) c_{12}(k,\alpha,l)=c_{21}(k,\alpha,l)\>,
\end{equation}
where
\begin{equation}\label{3.20}
\gamma(k,\alpha) =\frac{k}{k+\frac{eQ}{y_+}}.
\end{equation}
From (\ref{3.14})-(\ref{3.19}) we get also
\begin{equation}\label{3.21}
\gamma(k,\alpha) (|c_{12}(k,\alpha,l)|^2-|c_{11}(k,\alpha,l)|^2)=1,
\end{equation}
\begin{equation}\label{3.22}
\gamma ^{-1}(k,\alpha)(|c_{21}(k,\alpha,l)|^2-|c_{22}(k,\alpha,l)|^2)=1.
\end{equation}
These equations can be written in the form
\begin{equation}\label{3.23}
\frac{1}{\gamma(k,\alpha)}\frac{1}{|c_{12}(k,\alpha,l)|^2}+
|\frac{c_{11}(k,\alpha,l)}{c_{12}(k,\alpha,l)}|^2=1,
\end{equation}
\begin{equation}\label{3.24}
\gamma(k,\alpha)\frac{1}{|c_{21}(k,\alpha,l)|^2}+
|\frac{c_{22}(k,\alpha,l)}{c_{21}(k,\alpha,l)}|^2=1.
\end{equation}
     We introduce now  solutions $\Psi^+(x,k,\alpha,l)$,
$\Psi^-(x,k,\alpha,l)$ of the equation (\ref{1.1})
satisfying the following conditions
\begin{equation}\label{3.25}
\Psi^+(x,k,\alpha,l)=
\cases{e^{i(k+\frac{eQ}{y_+})x)}+
s_{12}(k,\alpha,l)e^{-i(k+\frac{eQ}{y_+})x}+o(1),&$x\to-\infty$,\cr
s_{11}(k,\alpha,l)w_{ieQ,\frac{1}{2}}(-2ikx)+o(1),&$x\to+\infty$,\cr}
\end{equation}
\begin{eqnarray}\label{3.26}
\Psi^-(x,k,\alpha,l)=
\qquad\qquad\qquad\qquad\qquad\qquad\qquad\qquad\qquad\qquad\qquad\nonumber\\
=\cases{s_{22}(k,\alpha,l)e^{-i(k+\frac{eQ}{y_+})x}
+o(1),&$x\to-\infty$,\cr
w_{-ieQ,\frac{1}{2}}(2ikx)+
s_{21}(k,\alpha,l)w_{ieQ,\frac{1}{2}}(-2ikx)+o(1),&$x\to+\infty$.\cr}
\end{eqnarray}
As a result, from (\ref{3.9}) we obtain
\begin{equation}\label{3.27}
\cases{\Psi^+(x,k,\alpha,l)=\overline{\psi^-(x,k,\alpha,l)}+
s_{12}(k,\alpha,l)\psi^-(x,k,\alpha,l)=\cr
s_{11}(x,k,\alpha,l)\psi^+(x,k,\alpha,l),\cr
\Psi^-(x,k,\alpha,l)=s_{22}(k,\alpha,l)\psi^-(x,k,\alpha,l)=\cr
\overline{\psi^+(x,k,\alpha,l)}+s_{21}(k,\alpha,l)\psi^+(x,k,\alpha,l).\cr}
\end{equation}
In the similar way, from (\ref{3.9}) we get
\begin{eqnarray}\label{3.28}
\Psi^-(x,k,\alpha,l)=\frac{\psi^-(x,k,\alpha,l)}{c_{12}(k,\alpha,l)},
\nonumber\\
s_{21}(k,\alpha,l)=\frac{c_{11}(k,\alpha,l)}{c_{12}(k,\alpha,l)},\qquad
s_{22}(k,\alpha,l)=\frac{1}{c_{12}(k,\alpha,l)},
\end{eqnarray}
and
\begin{eqnarray}\label{3.29}
\Psi^+(x,k,\alpha,l)=\frac{\psi^+(x,k,\alpha,l)}{c_{21}(k,\alpha,l)},
\nonumber\\
s_{12}(k,\alpha,l)=\frac{c_{22}(k,\alpha,l)}{c_{21}(k,\alpha,l)},\qquad
s_{11}(k,\alpha,l)=\frac{1}{c_{21}(k,\alpha,l)}.
\end{eqnarray}
     For example for the transition coefficient $s_{11}(k,\alpha,l)$
which stands in the expression (\ref{1.1a}) for the luminosity
$L(n,\alpha)$ we have formula

\begin{equation}              \label{3.32}
s_{11}(k,\alpha,l)=
2i(k+\frac{eQ}{y_+})/[\psi^-(x,k,\alpha,l),\psi^+(x,k,\alpha,l)]\>.
\end{equation}

According to (\ref{3.23}), (\ref{3.24}) we get the unitarity relations
for matrix $S=\{s_{ij}\}$,
\begin{equation}\label{3.30}
\gamma^{-1}(k,\alpha)|s_{22}(k,\alpha,l)|^2+|s_{21}(k,\alpha,l)|^2=1,
\end{equation}
\begin{equation}\label{3.31}
\gamma(k,\alpha) |s_{11}(k,\alpha,l)|^2+|s_{12}(k,\alpha,l)|^2=1.
\end{equation}

\section{Asymptotic properties of $S$-matrix}

     To find out asymptotic behavior of the elements of $S$-matrix as
$k\to\infty$ we should at first obtain the representation formulas
for Jost type functions $\psi^{\pm}(x,k,\alpha,l)$.
They are as follows
\begin{eqnarray} \label{4.1}
\psi^-(x,k,\alpha,l)=F^-(x)e^{-i(k+\frac{eQ}{y_+})x}+
\int^x_{-\infty}A^-(x,t)\psi^-(t,k,\alpha,l)dt,\qquad\>,
\end{eqnarray}
\noindent
\begin{equation}\label{4.2}
F^-(x)=\exp[i\int^x_{-\infty}v^-(x)dx]
\end{equation}
and
\begin{eqnarray} \label{4.3}
\psi^+(x,k,\alpha,l)=F^+(x)w_{ieQ,\frac{1}{2}}(-2ikx)+
\int_x^{+\infty}A^+(x,t)\psi^+(t,k,\alpha,l)dt,\qquad\>,
\end{eqnarray}
\begin{equation}\label{4.4}
F^+(x)=\exp[i\int_x^{+\infty}v^+(x)dx]\>.
\end{equation}
We have here
\begin{equation} \label{4.4a}
|\int_x^{\pm \infty}|A^{\pm}(x,t)|dt|<
C|\int_x^{\pm\infty}(|u^{\pm}(x)|+|(v^{\pm})'(x)|)dx|\>.
\end{equation}
    To get formulas (\ref{4.1})-(\ref{4.4}) one should replace
$\psi^{\pm}(x,k,\alpha,l)$ in the integral equation (\ref{3.3}), (\ref{3.7})
by its representations (\ref{4.1}), (\ref{4.3}) as usual (see Ref.\cite{J}).
Then we obtain the equations for $A^{\pm}(x,t)$ and $F^{\pm}(x)$. Solving
equations for $F^{\pm}(x)$ we get formulas (\ref{4.2}), (\ref{4.4}).

      From (\ref{4.1})-(\ref{4.4a}) it follows that at $k\to\infty$
\begin{eqnarray}\label{4.6}
\psi^-(x,k,\alpha,l)=e^{-ikx}
\exp\{i\int^1_{-\infty}v^-(x)dx-\frac{ieQ}{y_+}-
ieQ\int_1^x\frac{dx}{y(x)}\}+o(1)\>,
\end{eqnarray}
\begin{eqnarray}\label{4.7}
\psi^+(x,k,\alpha,l)=e^{ikx}\times\nonumber\\
\exp\{ieQ\ln2k+i\int^1_{-\infty}v^+(x)dx-
ieQ\int_1^x\frac{dx}{y(x)}\}+o(1)\>.
\end{eqnarray}
    It is not difficult to show that asymptotics (\ref{4.6}), (\ref{4.7})
can be differentiated. So, we have
\begin{eqnarray}\label{4.8}
(\psi^-)'(x,k,\alpha,l)=-ike^{-ikx}\times\nonumber\\
\exp\{i\int^1_{-\infty}v^-(x)dx-
\frac{ieQ}{y_+}-ieQ\int_1^x\frac{dx}{y(x)}\}[1+O(\frac{1}{k})]\>,
\end{eqnarray}
\begin{eqnarray}\label{4.9}
(\psi^+)'(x,k,\alpha,l)=e^{ikx}\times\nonumber\\
\exp\{ieQ\ln2k+i\int^1_{-\infty}v^+(x)dx-
ieQ\int_1^x\frac{dx}{y(x)}\}[1+O(\frac{1}{k})]\>.
\end{eqnarray}
From (\ref{4.6})-(\ref{4.9}) it follows the asymptotic for Wronskian
\begin{eqnarray}
[\psi^-(x,k,\alpha,l),\psi^+(x,k,\alpha,l)]=\nonumber\\
2ik\exp\{ieQ[\ln2k-1/y_+\}\exp(iV) [1+O(\frac{1}{k})]
\label{4.10}\>,
\end{eqnarray}
where
\begin{equation}
V=\int_{-\infty}^{\infty}v(x)dx,
v(x)=\cases{v^+(x),
&$x\geq 1$,\cr
v^-(x),
&$x<1$,\cr}
\label{4.11}\>.
\end{equation}
Therefore from (\ref{3.12}), (\ref{3.29}) we have
\begin{equation} \label{4.12}
s_{11}(k,\alpha,l)=
\exp\{-ieQ[\ln2k-1/y_+]\}\exp(-iV)[1+O(\frac{1}{k})]\>.
\end{equation}
In the same way we get asymptotics (see (\ref{3.11}), (\ref{3.28}))
\begin{equation} \label{4.13}
s_{22}(k,\alpha,l)=\exp\{-ieQ[\ln2k-1/y_+]\}
\exp(-iV)[1+O(\frac{1}{k})]
\end{equation}
and (see also (\ref{3.10}), (\ref{3.13}))
\begin{equation} \label{4.14}
s_{12}(k,\alpha,l)=O(\frac{1}{k})\>,\qquad
s_{21}(k,\alpha,l)= O(\frac{1}{k})\>.
\end{equation}
 So, we have come to the conclusion that the correct statement of the
scattering problem for the equation (\ref{1.1}) at $e\ne0$, $Q\ne0$, 
$\alpha\ne0$, should include the
conditions (\ref{3.25})-(\ref{3.26}) rather than the ones
(\ref{1.1b})--(\ref{1.1c}) accepted in the physical papers (see, e. g.,
Ref.\cite{NF86} and references cited therein).
It may be understood from the physical point of view since the own
(external) electric field of RN black hole is the Coulomb one. But
as is known from the ordinary quantum mechanics (see, e. g., Ref.\cite{LL})
the Coulomb potential always requires special treatment so long as it is the
long-range one.

\section{Concluding remarks}
We considered one of the scattering problems which encounter in the
4D black hole physics.
Other ones will emerge when
both the type of field and the type of black hole vary \cite{Gon97}.
In principle, each field type (scalar, spinorial etc.) poses
its own scattering problem
which also depends on the type of black hole. It should also be taken into
account that there can exist TICs for many fields on black holes and, as
a consequence, this can increase the number of possibilities. But, as was
mentioned early
in the paper, the elements of the corresponding $S$-matrices
are the important ingredients when calculating miscellaneous quantum effects
connected with black holes. We hope, therefore, to continue the strict study
of a number of the mentioned problems within the framework of our further
investigations.

\nonumsection{Acknowledgement}
\noindent
     The author is deeply thankful to Yu. Goncharov for drawing her
attention to these questions and for useful discussions. The work was
supported in part by the Russian Foundation for Basic Research (grant
No. 98-02-18244-a).

\nonumsection{References}

\end{document}